\documentclass[twocolumn,prl,floatfix,superscriptaddress,longbibliography]{revtex4-2}

\usepackage{amsmath}
\usepackage{amssymb}
\usepackage{graphicx}
\usepackage{hyperref}
\usepackage{float}
\usepackage[normalem]{ulem}

\makeatletter
\renewcommand{\sb}[1]{_{\mathrm{#1}}}

\makeatother

\usepackage{cleveref}
\crefname{figure}{Fig.}{Figs.} 

\usepackage{xcolor}
\usepackage{color}

\hypersetup{colorlinks=true,
    linkcolor={blue!80!black},
    citecolor={blue!80!black},
    urlcolor={blue!80!black}}

\begin{document}

\title{Continuous tuning of quantum critical behavior in hybrid Josephson junction arrays}

\author{E.~Fugl}
\affiliation{Niels Bohr Institute, University of Copenhagen, 2100 Copenhagen, Denmark}
\author{L.~L.~Olsen}
\affiliation{Niels Bohr Institute, University of Copenhagen, 2100 Copenhagen, Denmark}
\author{S.~Sasmal}
\affiliation{Niels Bohr Institute, University of Copenhagen, 2100 Copenhagen, Denmark}
\author{M.~Efthymiou-Tsironi}
\affiliation{Niels Bohr Institute, University of Copenhagen, 2100 Copenhagen, Denmark}
\affiliation{Department of Physics, Universitá del Salento, via Monteroni, 73100, Lecce, Italy}
\author{G.~Nagda}
\affiliation{Niels Bohr Institute, University of Copenhagen, 2100 Copenhagen, Denmark}
\author{F.~Krizek}
\affiliation{Niels Bohr Institute, University of Copenhagen, 2100 Copenhagen, Denmark}
\affiliation{Institute of Physics, Czech Academy of Sciences, 162 00 Prague, Czech Republic}
\affiliation{Department of Materials, Faculty of Nuclear Sciences and Physical Engineering, Czech Technical University in Prague, 11519 Prague, Czech Republic}
\author{L.~Casparis}
\affiliation{Niels Bohr Institute, University of Copenhagen, 2100 Copenhagen, Denmark}
\author{S.~Vaitiek\.{e}nas}
\affiliation{Niels Bohr Institute, University of Copenhagen, 2100 Copenhagen, Denmark}

\date{\today}

\begin{abstract} 

We report finite-temperature scaling of voltage-driven quantum phase transitions in hybrid Josephson junction arrays. A dual-gate geometry separately tunes interisland coupling and proximity-induced superconductivity, providing access to superconducting, insulating, and metallic transport regimes. The critical resistivity and scaling exponent evolve continuously with electrostatic tuning. Where the three regimes converge, the critical resistivity approaches $h/4e^2$, consistent with the self-dual limit, while the scaling exponent approaches the quantum-percolation estimate. Tuning toward stronger insulating behavior raises the critical resistivity and shifts the scaling exponent toward the classical-percolation value, while both parameters decrease toward the metallic regime. These results connect scaling regimes previously reported in different two-dimensional superconducting systems through continuous electrostatic control.

\end{abstract}

\maketitle

The superconductor-to-insulator transition (SIT) in two dimensions provides a paradigmatic example of a quantum phase transition, characterized by universal critical behavior~\cite{Wen_1990_Universal, Goldman_2010_Superconductor}. 
Within the dirty-boson framework, a self-dual SIT is predicted to exhibit a critical resistivity, $\rho\sb{C}$, equal to the resistance quantum of Cooper pairs, $\rho\sb{Q}=h/4e^2$~\cite{Fisher_1990_Presence}.
However, numerical studies have predicted deviations from this critical resistance~\cite{Cha_1991_Universal,Wallin_1994_DirtyBosons,Swanson_2014_Dynamical}, while experiments have reported a broad range of critical resistivities and scaling exponents~\cite{Kapitulnik_2019_Colloquium,Wang_2023_Quantum,Feldman_2025_Quantum,Rogachev_2025_Quantum}. 
Such variations have been attributed, among other factors, to disorder and coupling to a dissipative environment~\cite{Wagenblast_1997_Superconductor,Mason_1998_Dissipation}. 
Understanding what determines the critical behavior across different superconducting systems therefore remains an outstanding challenge.

In two-dimensional superconducting films, the SIT has conventionally been studied by varying disorder or applying a magnetic field~\cite{Fisher_1990_Quantum}. 
Disorder can be controlled through film thickness and growth conditions~\cite{Haviland_1989_Onset, Jaeger_1989_Onset}, while a magnetic field can drive the transition by suppressing superconductivity and promoting the localization of Cooper pairs~\cite{Yazdani_1995_Superconducting, Mason_2001_True, Aubin_2006_Magnetic,Biscaras_2013_Multiple, Breznay_2016_SelfDuality, Kronfeldner_2021_Multiple}. 
Studies of thin films have shown that increasing disorder can drive the critical resistivity $\rho\sb{C}$ toward $\rho\sb{Q}$~\cite{Steiner_2008_Approach}, whereas reduced disorder is associated with lower critical resistivities and weakly insulating transport near the transition~\cite{Breznay_2017_Superconductor}.
In thin films, however, the microscopic disorder landscape is largely fixed during fabrication, limiting the ability to systematically investigate its influence on critical behavior within a single device.

Josephson junction arrays (JJAs) offer a complementary platform to as-grown thin films, thanks to the flexible design of lattice geometries and controllable superconducting island couplings~\cite{Fazio_2001_Quantum}.
In conventional tunnel-junction arrays, the transition is governed by the competition between Josephson coupling and charging energy, as well as dissipation~\cite{Geerligs_1989_Charging, vanDerZant_1996_Quantum, Rimberg_1997_DissipationDrivenSIT, Takahide_2000_Superconductor}. 
Magnetic frustration provides another means of controlling vortex dynamics and modifying the critical behavior~\cite{vanDerZant_1992_FieldInduced, Chen_1995_Scaling, Chen_1996_Flux}. 
Recent advances in semiconductor--superconductor hybrids have enabled JJAs with gate-tunable junction transparencies, providing continuous control over the interisland Josephson coupling~\cite{Han_2014_Collapse, Bottcher_2018_Superconducting}. 
Such arrays have provided access to distinct quantum phases and their transitions through electrostatic gating~\cite{Bottcher_2018_Superconducting, Sasmal_2025_VoltageTuned, Leonard_2026_Radiometry}.
Spatially resolved measurements near these transitions have revealed the evolution of phase coherence and quantum fluctuations ~\cite{Wang_2026_Coherence, BishopVanHorn_2026_Imaging}.
These capabilities provide a means to systematically investigate how critical behavior evolves within a single device.

\begin{figure}[!t]
    \includegraphics[width=\linewidth]{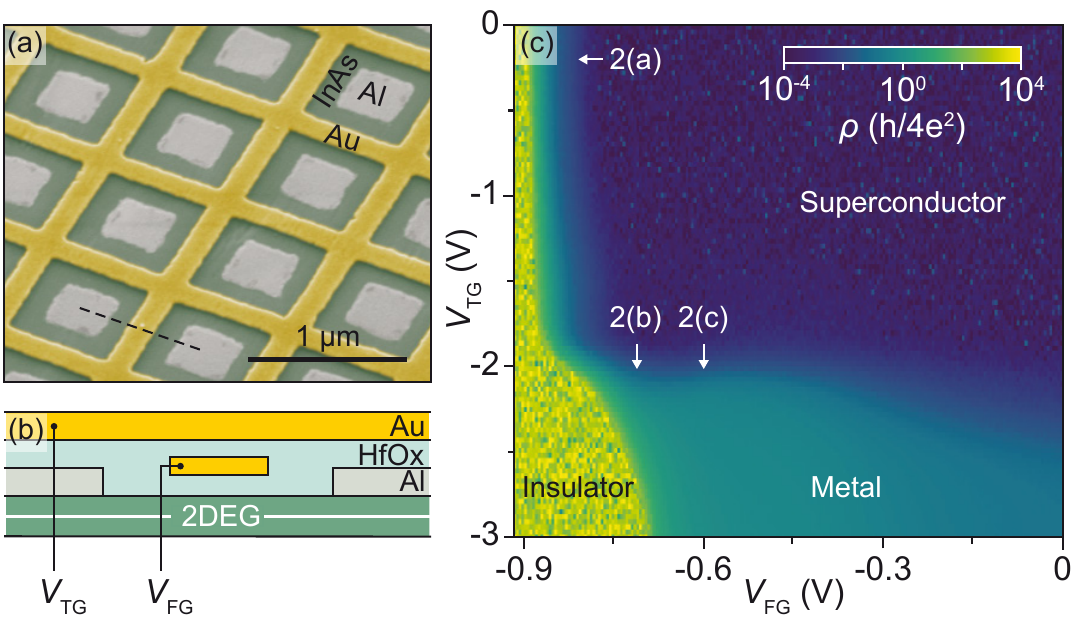}
    \caption{(a) Colorized scanning electron micrograph of a reference device, before depositing the top gate. The semiconductor is shown in green, Al islands in gray, and the frame gate in yellow. 
    (b) Schematic cross-section along the dashed black line in (a), illustrating the position of the two gate layers relative to the Josephson junctions. The frame-gate voltage ($V\sb{FG}$) controls the central part of the junctions, while the top-gate voltage ($V\sb{TG}$) tunes the semiconductor surrounding the islands.
    (c) Resistivity, $\rho$, of the array measured as a function of $V\sb{FG}$ and $V\sb{TG}$, showing superconducting, insulating, and metallic transport regimes.
    Arrows indicate the gate-voltage trajectories examined in Fig.~\ref{fig:2}.
    }
    \label{fig:1}
\end{figure}

Here, we investigate the evolution of critical behavior in square-lattice Josephson junction arrays based on hybrid semiconductor--superconductor heterostructures; see Fig.~\ref{fig:1}(a). 
The arrays are equipped with a dual-gate geometry that allows independent tuning of the interisland coupling and proximity-induced superconductivity in the surrounding semiconductor; see Fig.~\ref{fig:1}(b). 
This provides access to a phase diagram comprising superconducting, insulating, and metallic regimes; see Fig.~\ref{fig:1}(c). 
Using finite-temperature scaling analysis, we find that the critical resistivity, $\rho\sb{C}$, and scaling exponent, $\alpha$, evolve continuously within a single device. 
Near the point where the three regimes converge, $\rho\sb{C}$ approaches the Cooper-pair resistance quantum $\rho\sb{Q}$, while $\alpha$ approaches the value associated with quantum percolation. Depending on the direction of electrostatic tuning away from this point, $\rho\sb{C}$ increases above or decreases below $\rho\sb{Q}$, accompanied by scaling exponents approaching classical-percolation values or becoming progressively smaller toward the metallic regime. 
Our results thus connect critical behaviors previously observed in different superconducting systems through electrostatic control of a single array.

The arrays were fabricated in a Hall-bar geometry from an InAs two-dimensional electron gas (2DEG) proximitized by epitaxial Al~\cite{Shabani_2016_TwoDimensional, Cheah_2023_SelectiveArea, Zhang_2023_mobility}.
Each array consists of $498\times68$ square Al islands with 300~nm sides and 350~nm spacing between them, and is equipped with two gate electrodes [Fig.~\ref{fig:1}].
The first gate layer is a 150~nm wide metallic frame gate (FG) forming a grid around the islands; its voltage, $V\sb{FG}$, controls their Josephson coupling by depleting the semiconductor in the central regions of the junctions. 
The second layer is a top gate (TG) covering the entire array; its voltage, $V\sb{TG}$, tunes proximity-induced superconductivity in the semiconductor surrounding the islands.
Measurements were performed on two nominally identical devices, which exhibited consistent behavior. We focus on one device in the main text, with supporting data from the second presented in the Supplemental Material~\cite{Supplement}. 
The devices were measured using standard lock-in techniques, in a dilution refrigerator equipped with a vector magnet and a base temperature of 30~mK. 
Further fabrication and measurement details, including magnetic-field offsets and excitation-induced heating characterization, are given in the Supplemental Material~\cite{Supplement}.

We begin by mapping the resistivity, $\rho$, as a function of $V\sb{FG}$ and $V\sb{TG}$, revealing three distinct transport regimes, which we refer to as superconducting, insulating, and metallic; see Fig.~\ref{fig:1}(c). 
These regimes are distinguished by their low-temperature resistivities and temperature dependences. In the superconducting regime, the resistivity approaches zero upon cooling, with transport mediated by Josephson coupling between the Al islands. 
Depleting the semiconductor surrounding the islands suppresses proximity-induced superconductivity, leaving a square network of normally conducting semiconductor channels with a finite, weakly temperature-dependent resistivity of order $\rho\sb{Q}$.  
Further depletion of the channels between the islands drives the system into an insulating regime, characterized by a strongly increasing resistivity upon cooling. 
The two gates thus provide distinct tuning directions across the phase diagram, allowing both superconductor-to-insulator (SIT) and superconductor-to-metal (SMT) transitions to be accessed within the same array.

\begin{figure}[!t] 
    \includegraphics[width=\linewidth]{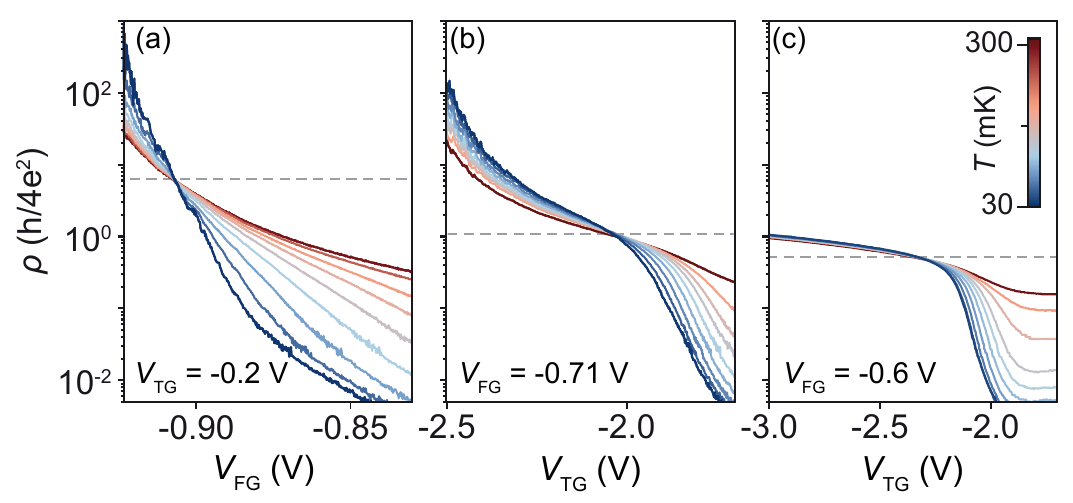}
    \caption{
    (a) Traces of $\rho$ as a function of $V\sb{FG}$ taken between 30 and 300~mK at fixed $V\sb{TG}=-0.2$~V, showing a crossing at $\rho\sb{C}\approx5.8\rho\sb{Q}$ with a pronounced separation of isotherms on either side, indicating a superconductor-to-insulator transition (SIT). 
    (b) Similar to (a), but varying $V\sb{TG}$ at fixed $V\sb{FG}=-0.71$~V, with a crossing near $\rho\sb{Q}$ and pronounced temperature dependence on either side, characteristic of another SIT. 
    (c) Same as (b) at $V\sb{FG}=-0.6$~V, where the crossing occurs around $0.5\rho\sb{Q}$, with a much weaker temperature dependence on the metallic side, indicating a superconductor-to-metal transition (SMT).
    }
    \label{fig:2}
\end{figure}

To investigate the transitions between these regimes, we examine three representative gate-voltage trajectories in the $V\sb{FG}$–$V\sb{TG}$ plane, marked by arrows in Fig.~\ref{fig:1}(c). 
Both gates weaken the Josephson coupling between Al islands, but their effects on the global transport behavior differ considerably. 
At fixed $V\sb{TG}$, applying a more negative $V\sb{FG}$ depletes the central regions of the junctions, progressively suppressing the interisland coupling. 
The resistivity traces measured for $V\sb{TG}=-0.2$~V at different temperatures cross around $V\sb{FG,C}=-0.91$~V, where $\rho$ is nearly temperature independent and defines a critical resistivity $\rho\sb{C}\approx6\rho\sb{Q}$; see Fig.~\ref{fig:2}(a). 
On the more negative side of this crossing, $\rho$ increases upon cooling, whereas on the opposite side it decreases, characteristic of an SIT.

Moving toward the point where the three transport regimes meet, we switch to tuning $V\sb{TG}$ at fixed $V\sb{FG}=-0.71$~V. Applied voltage to the top gate depletes the semiconductor surrounding the islands, weakening the superconducting proximity effect and thereby reducing the effective Josephson coupling between islands.
The isotherms again exhibit a crossing, in this case around $V\sb{TG,C}=-2.02$~V, with a reduced $\rho\sb{C}\approx\rho\sb{Q}$; see Fig.~\ref{fig:2}(b).
On either side of the crossing, the resistivity exhibits opposite temperature dependences, with an increasingly pronounced insulating behavior at more negative $V\sb{TG}$, indicating another SIT. 
At a less negative $V\sb{FG}$, however, the character of the top-gate-driven transition changes. 
At $V\sb{FG}=-0.6$ V, the semiconductor network remains conducting, and the resistivity on the more negative $V\sb{TG}$ side develops a much weaker temperature dependence, approaching a finite value at low temperatures. 
The isotherms still exhibit a crossing around $V\sb{TG,C}=-2.33$~V, now with $\rho\sb{C}<\rho\sb{Q}$ [Fig.~\ref{fig:2}(c)], characteristic of an SMT. 
These observations motivate a systematic finite-temperature scaling analysis to quantify how the critical behavior evolves across the phase diagram.

According to the scaling theory of the SIT, the correlation length, $\xi$, diverges near the critical gate voltage as $\xi\propto|V\sb{G}-V\sb{G,C}|^{-\nu}$, while the characteristic time scale follows $\tau\propto\xi^z$, where $\nu$ and $z$ are the correlation-length and dynamical critical exponents~\cite{Fisher_1990_Quantum}. These relations lead to the finite-temperature scaling form~\cite{Sondhi_1997_Continuous, Markovic_1999_SIT},
\begin{equation}\label{eq:scaling}
    \rho(T,V\sb{G})=\rho\sb{C} F\left(\vert V\sb{G}-V\sb{G,C}\vert\,T^{-1/z\nu}\right),
\end{equation}
where $F$ is a scaling function with separate branches on either side of the transition. The absolute value is used for a more direct comparison of the two regimes on either side of the transition. 
Experimentally, temperature scaling determines the inverse product $\alpha=1/(z\nu)$, which we extract from the power-law temperature dependence of the isotherm slopes around the crossing point. Further details of the analysis are provided in the Supplemental Material~\cite{Supplement}.

\begin{figure}[!t] 
    \includegraphics[width=\linewidth]{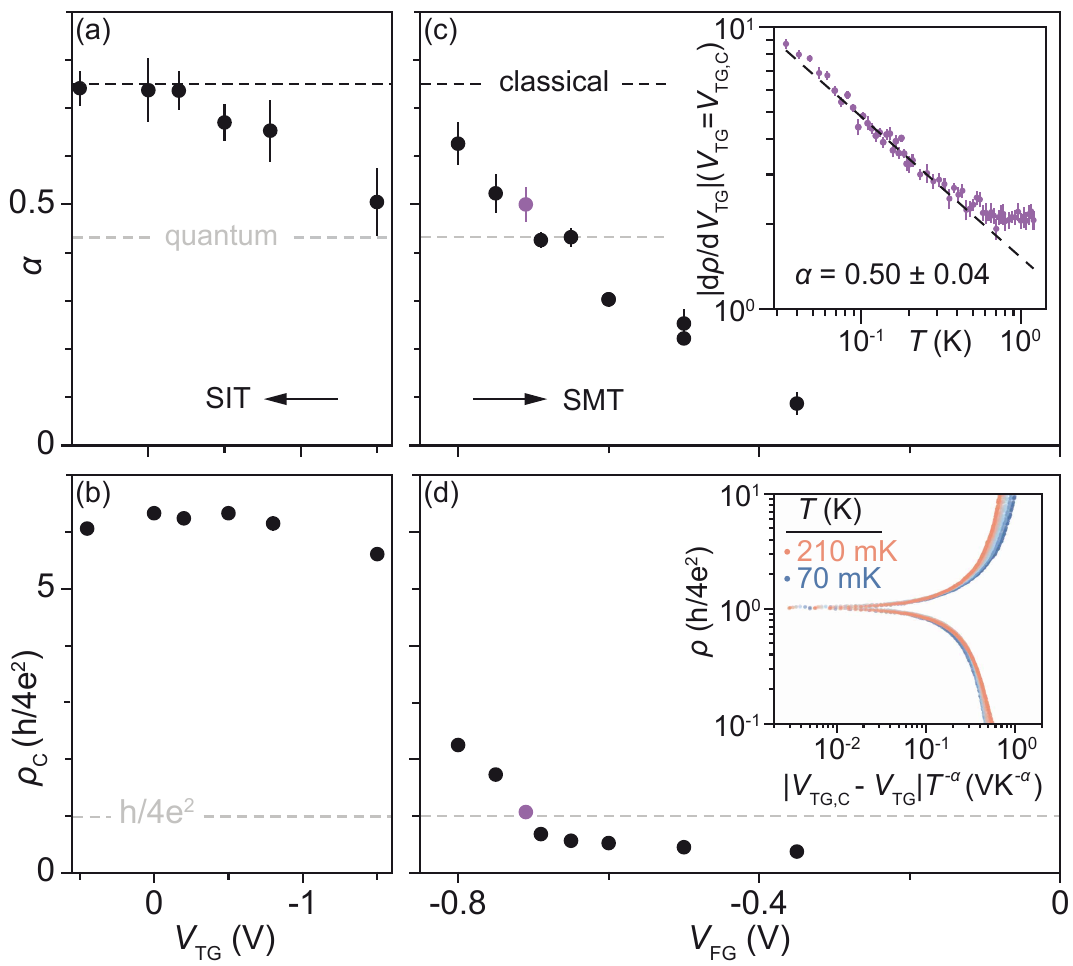}
    \caption{
    Evolution of the scaling exponent $\alpha$ and critical resistivity $\rho\sb{C}$ across the gate-voltage phase diagram. 
    (a) and (b) Critical parameters extracted from frame-gate-driven SITs at different fixed top-gate voltages, $V\sb{TG}$, showing a gradual decrease in both quantities toward the triple point. 
    (c) and (d) Corresponding parameters for top-gate-driven SITs and SMTs at different fixed frame-gate voltages, $V\sb{FG}$, showing a further decrease toward the metallic regime. Dashed horizontal lines indicate the classical- and quantum-percolation estimates for $\alpha$ and the Cooper-pair resistance quantum $\rho\sb{Q}=h/4e^2$. Insets show the power-law temperature dependence of the isotherm slopes, yielding $\alpha=0.50\pm0.04$ in (c), and the corresponding scaling collapse in (d), both for $V\sb{FG}=-0.71$~V.}
    \label{fig:3}
\end{figure}

We first apply this analysis to frame-gate-driven SITs at different fixed top-gate voltages. When the semiconductor regions surrounding the islands are highly populated, the extracted scaling exponent approaches the classical-percolation value $\alpha=3/4$ (assuming $z=1$)~\cite{Nijs_1979_Relation, Sondhi_1997_Continuous}, accompanied by a critical resistivity considerably larger than $\rho\sb{Q}$; see Figs.~\ref{fig:3}(a) and \ref{fig:3}(b). 
As $V\sb{TG}$ is decreased, both $\alpha$ and $\rho\sb{C}$ start to decrease, though the change is more pronounced for $\alpha$. 
In particular, $\alpha$ approaches the numerically estimated quantum-percolation value $\alpha\approx0.43$~\cite{Lee_1993_QPexponent, Dubi_2005_Unifying}, suggesting a change in the effective critical behavior as the system approaches the triple point where the three transport regimes converge.

For a direct comparison, we apply the same analysis to the complementary tuning direction, examining top-gate-driven transitions at different fixed frame-gate voltages. 
Around the triple point, both critical parameters approach theoretically significant values, with $\rho\sb{C}\approx\rho\sb{Q}$ and $\alpha$ close to the quantum-percolation estimate; see Figs.~\ref{fig:3}(c) and \ref{fig:3}(d). 
For the representative SIT at $V\sb{FG}=-0.71$~V, also shown in Fig.~\ref{fig:2}(b), the isotherm slopes exhibit a power-law temperature dependence with $\alpha=0.50\pm0.04$, yielding a collapse of the resistivity traces onto two scaling branches [insets of Figs.~\ref{fig:3}(c) and \ref{fig:3}(d)]. 
As $V\sb{FG}$ is increased toward zero, both $\rho\sb{C}$ and $\alpha$ decrease continuously.
We note that in this regime the array displays SMT behavior.
Nevertheless, we extend the conventional SIT scaling analysis into the metallic regime to characterize the continued evolution of the crossing behavior, treating the extracted parameters as effective rather than necessarily asymptotic critical values.
Eventually, the temperature dependence becomes sufficiently weak that a well-defined crossing point can no longer be identified.

The continuous evolution of the critical parameters provides insight into the nature of the SIT under different electrostatic conditions. 
Within the dirty-boson framework, the critical resistivity $\rho\sb{C}\approx\rho\sb{Q}$ observed near the triple point is consistent with the self-dual limit~\cite{Fisher_1990_Presence}, while the simultaneous approach of $\alpha$ toward the quantum-percolation value suggests a distinct regime of critical behavior. 
Microscopically, this may correspond to a regime with a single occupied electronic subband and reduced scattering in the semiconductor junctions. 
Shallow InAs quantum wells exhibit a characteristic nonmonotonic mobility dependence on carrier density, associated in part with higher-subband occupation and scattering~\cite{Cheah_2023_SelectiveArea,Zhang_2023_mobility}, with a similar mobility maximum observed in our devices in the normal state (see Supplemental Material~\cite{Supplement}).

Increasing $V\sb{TG}$ away from the triple point populates the semiconductor surrounding the islands, introducing additional inter-subband and interface scattering. 
The accompanying increase in $\rho\sb{C}$ is thus consistent with theoretical predictions that disorder can enhance the critical resistivity~\cite{Cha_1991_Universal}. 
The simultaneous increase in $\alpha$ is consistent with stronger localization effects on the insulating side due to the increased scattering. 
Increasing $V\sb{FG}$, in contrast, populates the central regions of the junctions while their proximity coupling to the superconducting islands remains relatively weak. 
In this metallic regime, the decrease in $\alpha$ reflects the progressively weaker temperature dependence of the resistivity. 
The simultaneous reduction of $\rho\sb{C}$ below $\rho\sb{Q}$, however, suggests an increasing contribution from normal-state transport, consistent with earlier observations and theoretical descriptions of superconductor-to-metal transitions in highly conducting systems~\cite{Steiner_2008_Approach, Spivak_2008_Theory, Gupta_2024_SMT}. 
To seek further clues about the transport mechanisms in this regime, we examine how the magnetoresistance of the semiconductor network evolves with $V\sb{FG}$.

\begin{figure}[!t] 
    \includegraphics[width=\linewidth]{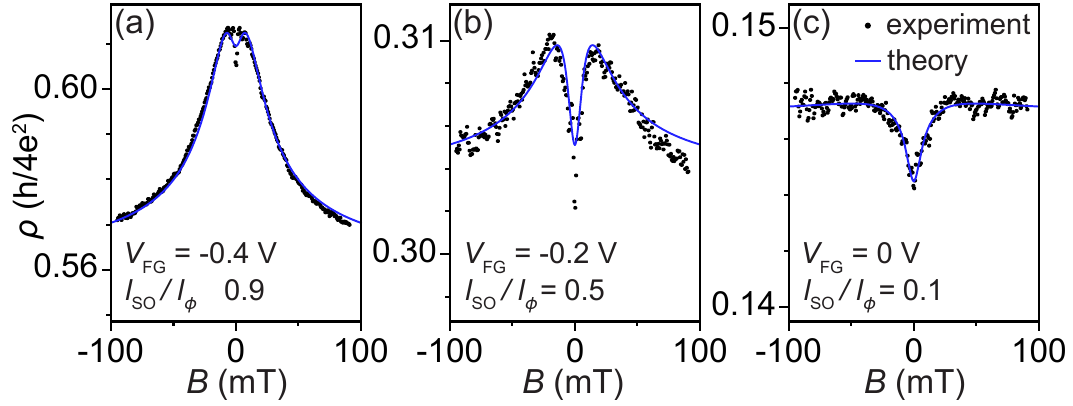}
    \caption{Magnetoresistivity of the semiconductor network measured at $V\sb{TG}=-4$~V and 150~mK for (a) $V\sb{FG}=-0.4$~V, (b) $-0.2$~V, and (c) $0$~V.
    Increasing $V\sb{FG}$ drives a crossover from weak-localization (WL) to weak-antilocalization (WAL) dominated transport. 
    Black points represent experimental data, and blue lines are fits to a quasi-one-dimensional interference model from Ref.~\cite{Hansen_2005_Spin}. 
    The extracted ratios of spin-orbit and phase-coherence lengths, $l\sb{SO}/l\sb{\phi}$, are indicated in each panel.
    }
    \label{fig:4}
\end{figure}

We investigate magnetotransport in the metallic regime, at $V\sb{TG}=-4$~V, where the superconducting islands are effectively decoupled and transport is mediated by a square network of semiconductor channels. 
Starting from negative frame-gate voltages, the magnetoresistivity exhibits a broad weak-localization (WL) peak around zero magnetic field [Fig.~\ref{fig:4}(a)]. 
Increasing $V\sb{FG}$ progressively populates the semiconductor network, suppressing the WL contribution and revealing an increasingly pronounced weak-antilocalization (WAL) dip [Fig.~\ref{fig:4}(b)]. 
At $V\sb{FG}=0$, the magnetoresistance is dominated by WAL, with no discernible WL peak within the measured range [Fig.~\ref{fig:4}(c)]. 
This evolution is consistent with the weakening insulating behavior as the semiconductor network becomes more conducting.

To characterize this crossover, we fit the magnetoresistivity using a quasi-one-dimensional interference model~\cite{Hansen_2005_Spin}, which includes the spin-orbit length $l\sb{SO}$ and phase-coherence length $l\sb{\phi}$ among its fitting parameters. 
The use of this model is motivated by the narrow semiconductor channels, whose transverse width, set by the frame-gate geometry, is much smaller than the fitted phase-coherence length ($l\sb{\phi}\gtrsim1~\mu$m) over the gate-voltage range considered here.
Independent Hall measurements on the same heterostructure showed a mean free path on the order of 100~nm~\cite{Sasmal_2025_VoltageTuned}.
We employ the dirty-limit approximation and focus on the relative evolution of $l\sb{SO}/l\sb{\phi}$ rather than the absolute values of the fitted length scales (see Supplemental Material~\cite{Supplement}). 
The extracted ratio decreases progressively from approximately 0.9 at $V\sb{FG}=-0.4$~V, through 0.5 at $-0.2$~V, to 0.1 at $0$~V, consistent with the observed evolution from WL- to WAL-dominated transport. 
The enhancement of WAL at higher carrier densities (less negative $V\sb{FG}$) is consistent with observations of increased spin-orbit splitting in shallow InAs quantum wells~\cite{Witt_2023_Spin}. 
The growing WAL contribution introduces a negative correction to resistivity, counteracting localization and offering a natural mechanism for the weakening insulating behavior and disappearance of the crossing points as $V\sb{FG}$ approaches zero.

In summary, we have studied hybrid Josephson junction arrays with a dual-gate geometry that allows \textit{in-situ} tuning between superconducting, insulating, and metallic transport regimes.
Using electrostatic control, we mapped the continuous evolution of the critical resistivity and scaling exponent characterizing the low-temperature transitions between these regimes. 
Near the point where the three regimes converge, the critical resistivity approaches $h/4e^2$, consistent with the self-dual limit, while the scaling exponent approaches the value associated with quantum percolation. 
We associate the observed evolution with gate-controlled changes in scattering and localization in the semiconductor network. 
Independent control of these electronic properties provides an experimental route to identifying the microscopic factors that govern quantum critical behavior and its variation across superconducting systems.
\\

\textit{Acknowledgments}---We thank S.~Upadhyay for assistance with the device fabrication and J.~Fugl, S.~Kivelson, K.~Lindholmer, C.~M.~Marcus, and A.~Padley for discussions. We acknowledge support from research grants (Projects No. 43951, No. 53097, and No. 50334) from VILLUM FONDEN, the Danish National Research Foundation, the European Research Council (Grant Agreement No. 856526), the MEYS grant (Project No. LM202351) and Lumina Quaeruntur fellowship (Project No. LQ100102602).\\

\bibliography{Literature.bib}

\end{document}


\title{Supplemental Material:\\Continuous tuning of quantum critical behavior in hybrid Josephson junction arrays}

\author{E.~Fugl}
\affiliation{Niels Bohr Institute, University of Copenhagen, 2100 Copenhagen, Denmark}
\author{L.~L.~Olsen}
\affiliation{Niels Bohr Institute, University of Copenhagen, 2100 Copenhagen, Denmark}
\author{S.~Sasmal}
\affiliation{Niels Bohr Institute, University of Copenhagen, 2100 Copenhagen, Denmark}
\author{M.~Efthymiou-Tsironi}
\affiliation{Niels Bohr Institute, University of Copenhagen, 2100 Copenhagen, Denmark}
\affiliation{Department of Physics, Universitá del Salento, via Monteroni, 73100, Lecce, Italy}
\author{G.~Nagda}
\affiliation{Niels Bohr Institute, University of Copenhagen, 2100 Copenhagen, Denmark}
\author{F.~Krizek}
\affiliation{Niels Bohr Institute, University of Copenhagen, 2100 Copenhagen, Denmark}
\affiliation{Institute of Physics, Czech Academy of Sciences, 162 00 Prague, Czech Republic}
\affiliation{Department of Materials, Faculty of Nuclear Sciences and Physical Engineering, Czech Technical University in Prague, 11519 Prague, Czech Republic}
\author{L.~Casparis}
\affiliation{Niels Bohr Institute, University of Copenhagen, 2100 Copenhagen, Denmark}
\author{S.~Vaitiek\.{e}nas}
\affiliation{Niels Bohr Institute, University of Copenhagen, 2100 Copenhagen, Denmark}

\date{\today}

\maketitle

\section{Sample Preparation}

The devices were fabricated in a Hall-bar geometry from semiconductor–superconductor heterostructures similar to those described in Refs.~\cite{Shabani_2016_TwoDimensional, Cheah_2023_SelectiveArea, Zhang_2023_mobility}.
The material was grown by molecular beam epitaxy (MBE) on a semi-insulating Fe-doped InP substrate. 
The heterostructure consists of an 8.5~nm thick InAs quantum well between 6.3 and 13.4~nm In$_{0.75}$Ga$_{0.25}$As bottom and top barriers, respectively, covered by a nominally 12~nm thick Al layer grown and subsequently oxidized \textit{in situ}.

Arrays of square Al islands, with 300-nm side lengths and 350-nm spacing between them, were defined by electron-beam lithography using adhesion promoter (Allresist, AR 300-80~new) and A4 PMMA resist. 
The Al was selectively removed by wet etching at 50$^\circ$~C for 5~s (Transene, Aluminum Etchant Type D). 
The devices were then patterned into Hall bars and electrically isolated using a wet etch (220:55:3:3 H$_2$O:C$_6$H$_8$O$_7$:H$_3$PO$_4$:H$_2$O$_2$). 
Normal-metal Ti/Au (5/100 nm) ohmic contacts were deposited (AJA International Inc., Orion) using A4 PMMA resist.
The sample was covered with a 15~nm HfO$\sb{X}$ gate dielectric deposited by atomic layer deposition (Veeco, Savannah S100). 
The frame gate was fabricated by lithographically defining a Ti/Au (5/20~nm) metallic grid on top of the mesa, followed by a second lithography step to connect the grid to the bond pads using thicker Ti/Au (15/380~nm) leads. 
A second 15~nm HfO$\sb{X}$ dielectric layer was then deposited, followed by a Ti/Au (15/380~nm) top gate covering the array.

\section{Measurement setup}
\begin{figure}[t]
    \includegraphics[width=\linewidth]{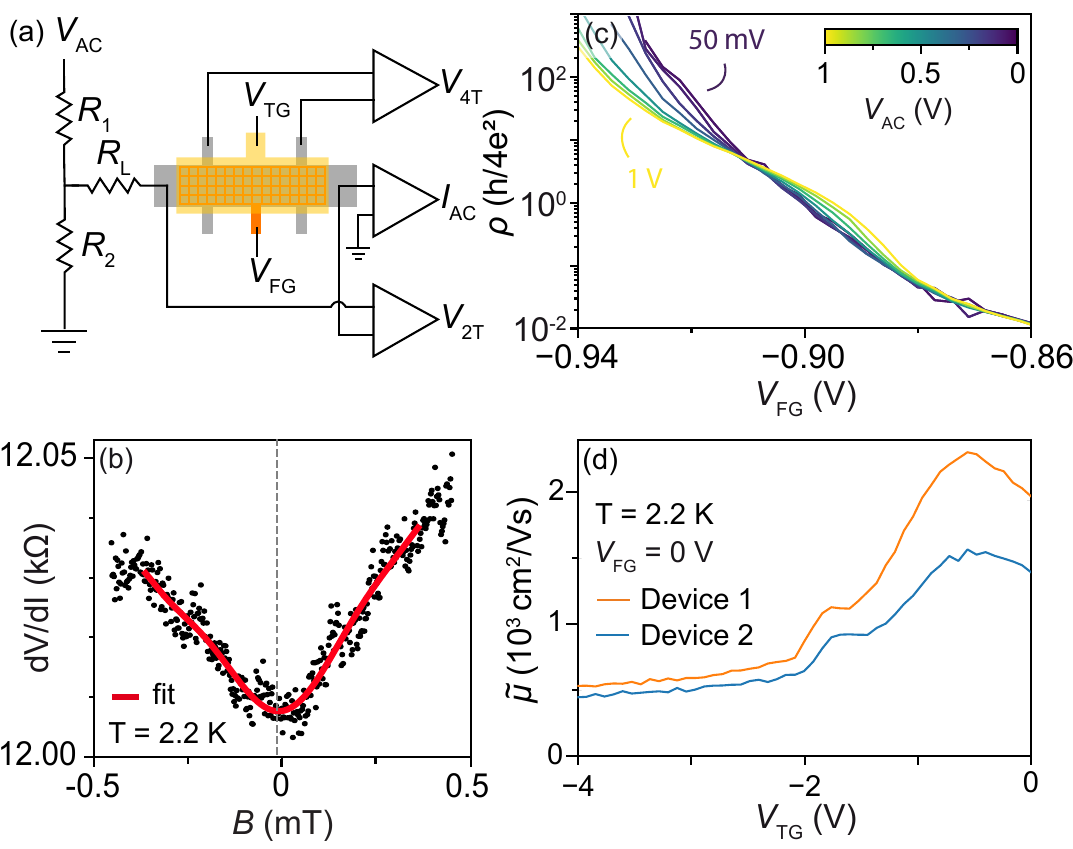}
    \caption{(a) Simplified circuit diagram of the ac current bias and voltage readout, including the quasi-two- and four-terminal probes. 
    (b) Differential resistance as a function of perpendicular magnetic field at $T=2.2$~K, above the superconducting transition temperature ($T\sb{C}=1.8$~K). The red curve is a polynomial fit to the weak-antilocalization (WAL) dip, used to determine the residual magnetic-field offset. 
    (c) Resistivity, $\rho$, across the frame-gate-driven SIT at $V\sb{TG}=0$ for different ac source excitation amplitudes, $V\sb{AC}$.
    (d) Effective mobility, $\tilde{\mu}$, as a function of $V\sb{TG}$ at fixed $V\sb{FG}=0$, extracted from normal-state Hall measurements at $T=2.2$~K for both devices.}
    \label{fig:s1}
\end{figure}

The devices were measured in a cryogen-free dilution refrigerator (Oxford Instruments, Triton 400) equipped with a three-axis (1, 1, 6)~T vector magnet and a base temperature of 30~mK. 
The temperatures reported throughout the manuscript correspond to the average mixing-chamber temperature recorded using a commercial RuO$_2$ sensor.
Electrical measurements were performed using standard ac lock-in techniques. Differential voltages were measured using voltage preamplifiers (Stanford Research Systems, SR560) and lock-in amplifiers (Stanford Research Systems, SR830). 
The ac excitation, $V\sb{AC}$, was generated by a separate lock-in amplifier (Stanford Research Systems, SR860) and applied through a home-built voltage divider ($R\sb{1}/R\sb{2}=10^{-3}$) and a load resistor $R\sb{L}=1$~M$\Omega$, in addition to roughly $1.6$~k$\Omega$ filter (line) resistances. The ac current was measured using a current-to-voltage converter (Basel Precision Instruments, SP983) connected to the same lock-in amplifier (SR860). A schematic of the measurement circuit is shown in Fig.~\ref{fig:s1}(a).

The sheet resistivity was determined using a \mbox{(quasi-)two-}terminal configuration, as some of the voltage side-probes become depleted by the frame gate before reaching the SIT. 
We have validated this approach against four-terminal measurements on the same device platform, showing consistent resistivities after correcting for residual contact and line resistances~\cite{Sasmal_2025_VoltageTuned}. 
The sheet resistivity is given by 
\begin{equation}
    \rho=(R\sb{2T}-R\sb{contact}-R\sb{line})\,W\sb{HB}/L\sb{HB}\,,
\end{equation}
\noindent where $W\sb{HB}=50~\mu$m and $L\sb{HB}=350~\mu$m are the Hall-bar width and length, respectively. 
The combined contact and line resistance was determined from the residual resistance measured in the superconducting regime. 
For device 1, separate bonding connections for current bias and voltage sensing yielded a residual resistance of 5.4~$\Omega$, attributed primarily to the contacts. 
For device 2, a shared connection was used for current bias and voltage sensing, resulting in a larger residual resistance of 911~$\Omega$, including contributions from the measurement lines. 

\section{Setup optimization}
To minimize the effects of residual magnetic field, we calibrated the field offset by warming the mixing chamber to 2.2~K, above the superconducting transition temperature of the Al islands, $T\sb{C}=1.8$~K, thereby releasing trapped vortices. 
In this normal-state regime, the differential resistance exhibits a weak-antilocalization (WAL) dip around zero perpendicular magnetic field. 
We identify the effective zero-field position from the minimum of a ninth-order polynomial fit to the magnetoresistance measured over a range of $\pm0.5$~mT [Fig.~\ref{fig:s1}(b)]. 
The applied magnetic field was then adjusted to compensate for the offset before cooling the device to base temperature.

To assess excitation-induced heating, we measured the frame-gate-driven SIT at $V\sb{TG}=0$ for source excitation amplitudes, $V\sb{AC}$, ranging from 50~mV to 1~V [Fig.~\ref{fig:s1}(c)], corresponding to nominal ac currents of approximately 50~pA to 1~nA. 
Increasing the excitation progressively modifies the resistivity traces, particularly on the insulating side of the transition, consistent with Joule heating. 
For SIT measurements, we therefore select $V\sb{AC}=100$~mV, below which the traces show no appreciable changes, even at resistivities considerably exceeding $\rho\sb{Q}$. 
For SMT measurements, where the resistivity remains substantially lower, the traces are largely unaffected by higher excitations, allowing us to increase $V\sb{AC}$ to 500~mV to further improve the signal-to-noise ratio. 

\section{Characterization}
To characterize normal-state transport in the semiconductor network, we measured the longitudinal and Hall resistivities, $\rho\sb{xx}$ and $\rho\sb{xy}$, at $T=2.2$~K, above the superconducting transition temperature. 
The measurements were performed as a function of $V\sb{TG}$ at fixed $V\sb{FG}=0$. Because the dual-gate geometry creates regions with different electrostatic environments, we refer to the extracted mobility as an effective mobility, $\tilde{\mu}$, and use it to characterize the gate dependence rather than the absolute mobility of the underlying 2DEG.

Both devices exhibit a nonmonotonic dependence of $\tilde{\mu}$ on $V\sb{TG}$, with two mobility maxima [Fig.~\ref{fig:s1}(d)]. 
We associate these features with the distinct semiconductor regions primarily controlled by the top and frame gates, respectively. 
Although $V\sb{FG}$ is fixed, the semiconductor beneath the frame gate remains sensitive to $V\sb{TG}$ through fringing fields. 
We note that the mobility maximum near $V\sb{TG}=-2$~V occurs close to the top-gate voltage of the triple point discussed in the main text.

\section{Finite-temperature scaling analysis}
\begin{figure}[t]
    \includegraphics[width=\linewidth]{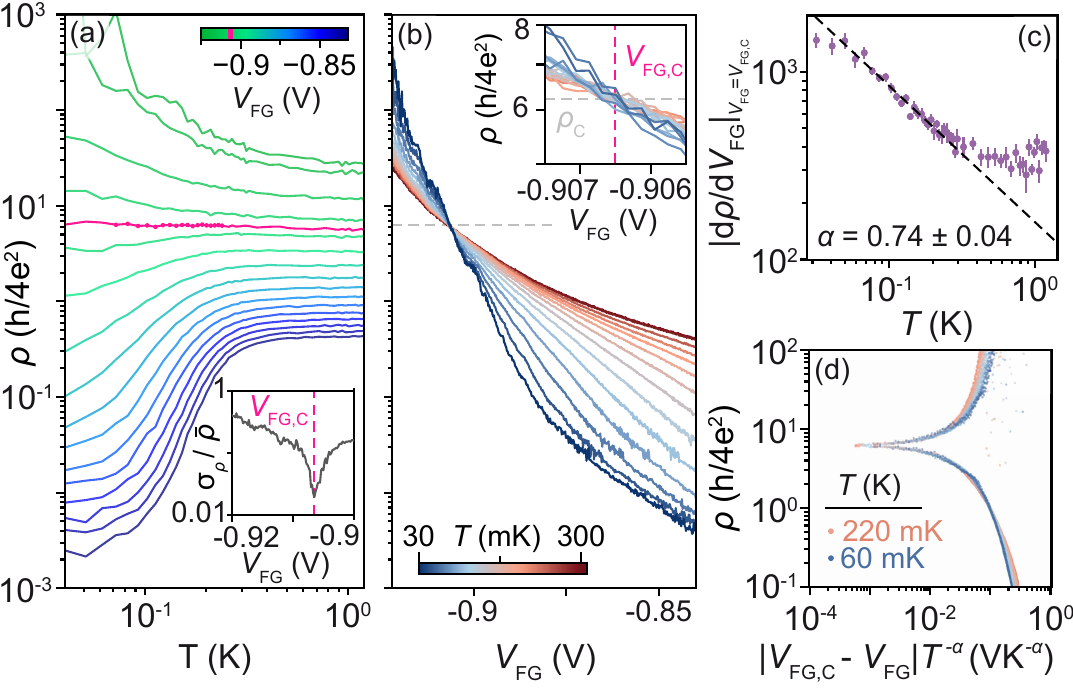}
    \caption{
    Finite-temperature scaling analysis of a frame-gate-driven SIT at a fixed top-gate voltage, $V\sb{TG}=-0.2$~V. 
    (a) Resistivity, $\rho$, as a function of temperature, $T$, for different frame-gate voltages, $V\sb{FG}$. The inset shows the normalized standard deviation, $\sigma\sb{\rho}/\overline{\rho}$, used to identify the critical gate voltage, $V\sb{FG,C}$. 
    (b) Resistivity isotherms as a function of $V\sb{FG}$. The inset highlights the crossing region, with the extracted $V\sb{FG,C}$ and critical resistivity, $\rho\sb{C}$. 
    (c) Temperature dependence of the isotherm slopes at $V\sb{FG,C}$, yielding the scaling exponent $\alpha=0.74\pm0.04$. 
    (d) Two-branch scaling collapse of the isotherms using the extracted critical parameters.
    }
    \label{fig:s2}
\end{figure}

\begin{figure}[t]
    \includegraphics[width=\linewidth]{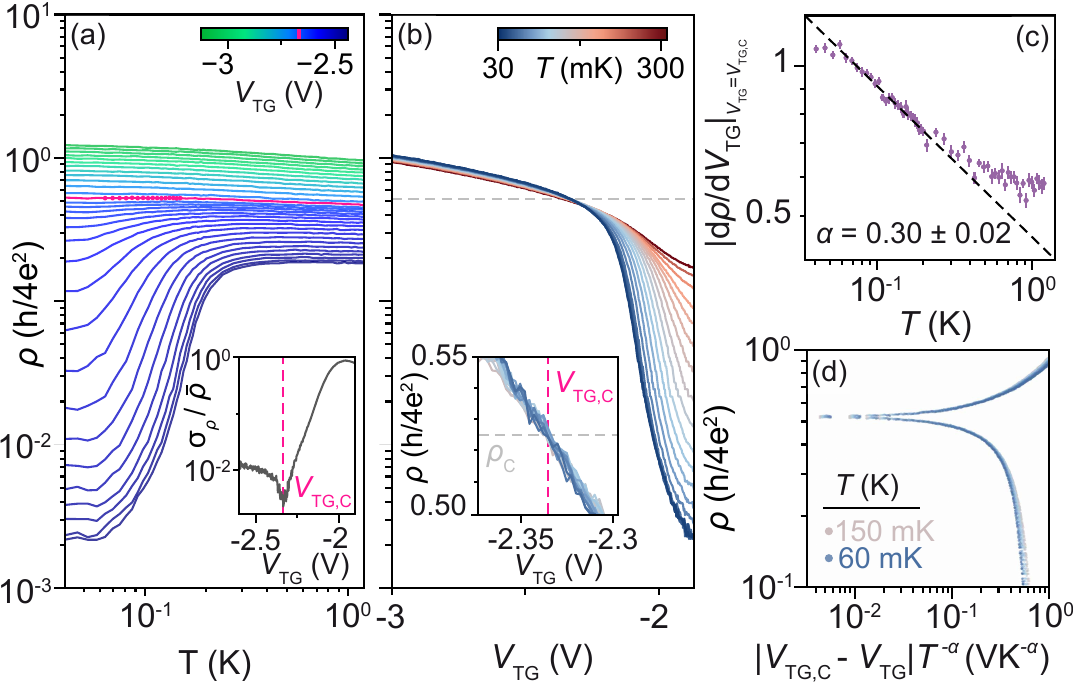}
    \caption{
    Finite-temperature scaling analysis of a top-gate-driven SMT at a fixed frame-gate voltage, $V\sb{FG}=-0.6$~V. 
    (a) Resistivity, $\rho$, as a function of temperature, $T$, for different top-gate voltages, $V\sb{TG}$. The inset shows the normalized standard deviation, $\sigma\sb{\rho}/\overline{\rho}$, used to identify the critical gate voltage $V\sb{TG,C}$. 
    (b) Resistivity isotherms as a function of $V\sb{TG}$. The inset highlights the crossing region, with the extracted $V\sb{TG,C}$ and critical resistivity, $\rho\sb{C}$. 
    (c) Temperature dependence of the isotherm slopes at $V\sb{TG,C}$, yielding the scaling exponent $\alpha=0.30\pm0.02$. 
    (d) Two-branch scaling collapse of the isotherms using the extracted critical parameters.
    }
    \label{fig:s3}
\end{figure}

We characterize the gate-driven transitions using the temperature dependence of the resistivity, $\rho(T,V\sb{G})$, where $V\sb{G}$ denotes the gate voltage driving the transition.
We focus on transitions exhibiting a well-defined crossing of the resistivity isotherms, where the resistivity becomes approximately temperature independent and defines the critical resistivity, $\rho\sb{C}$.
Representative examples are shown in Fig.~\ref{fig:s2} for a frame-gate-driven SIT and Fig.~\ref{fig:s3} for a top-gate-driven SMT.
As discussed in the main text, we extend the conventional SIT scaling analysis phenomenologically to SMTs exhibiting similar crossing behavior.

To identify the critical gate voltage, $V\sb{G,C}$, we calculate the normalized standard deviation, $\sigma\sb{\rho}/\overline{\rho}$, up to 300~mK for each gate voltage, where $\sigma\sb{\rho}$ denotes the standard deviation and $\overline{\rho}$ is the mean resistivity.
A pronounced minimum of this quantity confirms that there is a unique crossing point and identifies $V\sb{G,C}$, where the resistivity is least temperature dependent, corresponding to the crossing point of the isotherms and defining $\rho\sb{C}$ [insets of Figs.~\ref{fig:s2}(a) and \ref{fig:s3}(a)].

We extract the scaling exponent $\alpha=1/z\nu$ from the temperature dependence of the isotherm slopes near the crossing point. 
According to the finite-temperature scaling relation, $\rho(T,V\sb{G})=\rho\sb{C}\,F(|V\sb{G}-V\sb{G,C}|\,T^{-\alpha})$, with separate branches on opposite sides of the transition, we get

\begin{equation}\label{eq:alpha}
    \log\left|\partial\rho(T,V\sb{G})/\partial V\sb{G}\right|_{V\sb{G}=V\sb{G,C}}=\mathrm{const}-\alpha\,\log T\,.
\end{equation}

\noindent For each temperature, we obtain the derivative from a linear fit to the resistivity over a narrow gate-voltage range around $V\sb{G,C}$ [insets of Figs.~\ref{fig:s2}(b) and \ref{fig:s3}(b)].
The resulting slopes follow a power-law temperature dependence over a finite range, roughly between 30 and 300~mK. 
Linear fits to Eq.~\eqref{eq:alpha} on a logarithmic scale yield $\alpha=0.74\pm0.04$ for the representative SIT [Fig.~\ref{fig:s2}(c)] and $\alpha=0.30\pm0.02$ for the SMT [Fig.~\ref{fig:s3}(c)], with the uncertainties representing the fitting errors.
Using the extracted critical parameters, we replot the isotherms as a function of $|V\sb{G}-V\sb{G,C}|\,T^{-\alpha}$, obtaining a collapse onto two branches corresponding to opposite sides of the transition [Figs.~\ref{fig:s2}(d) and \ref{fig:s3}(d)].

\begin{figure}[t]
    \includegraphics[width=\linewidth]{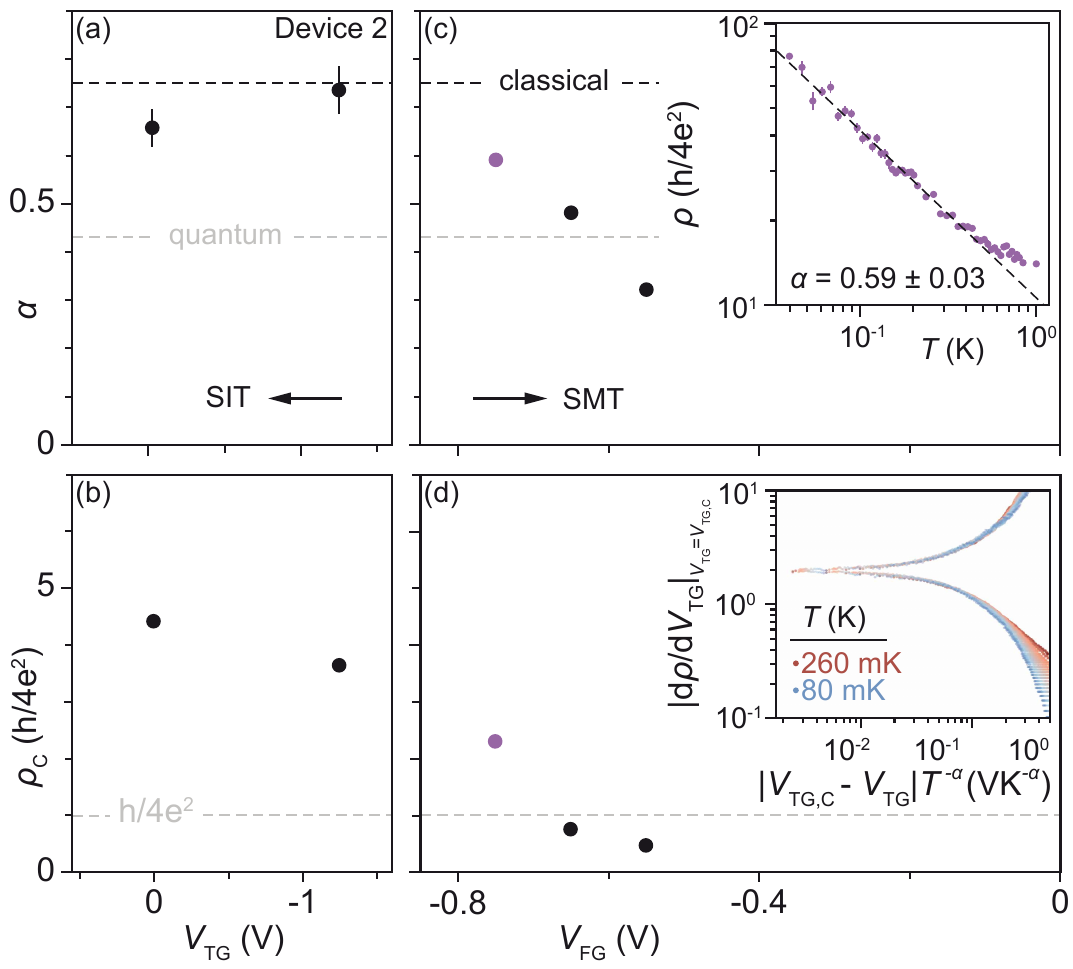}
    \caption{
    Evolution of the scaling exponent $\alpha$ and critical resistivity $\rho\sb{C}$ in Device 2. 
    (a) and (b) Critical parameters extracted from frame-gate-driven SITs at different fixed top-gate voltages $V\sb{TG}$. 
    (c) and (d) Corresponding parameters for top-gate-driven SITs and SMTs at different fixed frame-gate voltages $V\sb{FG}$, showing a decrease toward the metallic regime. 
    Dashed horizontal lines indicate the classical- and quantum-percolation estimates for $\alpha$ and the Cooper-pair resistance quantum $\rho\sb{Q}=h/4e^2$. Insets show the power-law temperature dependence of the isotherm slopes, yielding $\alpha=0.59\pm0.03$ in (c), and the corresponding scaling collapse in (d), both for $V\sb{FG}=-0.75$~V.
    }
    \label{fig:s4}
\end{figure}

\section{Additional device}

To assess the reproducibility of the observed critical behavior, we performed analogous measurements on a second device with nominally identical geometry. 
We followed frame- and top-gate trajectories over gate-voltage ranges comparable to those investigated in the main device. 
The extracted $\rho\sb{C}$ and $\alpha$ exhibit qualitatively the same gate-dependent trends [Fig.~\ref{fig:s4}], including a progressive decrease toward the metallic regime. 
For a representative top-gate-driven SIT at $V\sb{FG}=-0.75$~V, the scaling analysis yields $\alpha=0.59\pm0.03$ [inset of Fig.~\ref{fig:s4}(c)], with the corresponding isotherms exhibiting a two-branch scaling collapse [inset of Fig.~\ref{fig:s4}(d)]. These measurements demonstrate that the gate-dependent evolution of the critical parameters is reproducible across nominally identical arrays.

\section{Measuring localization effects}

At sufficiently negative top-gate voltage ($V\sb{TG}\lesssim-3$~V), the semiconductor regions surrounding the Al islands are strongly depleted, suppressing proximity-induced superconductivity and effectively decoupling the islands. 
Transport is then mediated by a network of gate-defined semiconductor channels, whose carrier density is controlled by $V\sb{FG}$. 
As discussed in the main text, increasing $V\sb{FG}$ progressively suppresses the broad weak-localization (WL) peak, revealing an increasingly pronounced weak-antilocalization (WAL) dip.

To characterize this evolution, we study the magnetotransport at $V\sb{TG}=-4$~V and fit it using a quasi-one-dimensional interference model~\cite{Hansen_2005_Spin}. 
The approximation is motivated by the narrow semiconductor channels, whose effective width is set by the frame-gate geometry and is smaller than the phase-coherence length. 
We employ the dirty-limit expression, derived for an elastic mean free path $l\sb{e}$ much smaller than the channel width $W$. 
Independent measurements on the same heterostructure found maximal $l\sb{e}\approx180$~nm at the mobility peak~\cite{Sasmal_2025_VoltageTuned}. 
Since $V\sb{TG}=-4$~V lies beyond the observed mobility maxima [Fig.~\ref{fig:s1}(d)], we expect a shorter mean free path in this regime, motivating the use of the dirty-limit approximation.
Taking the network geometry into consideration, we express the interference correction in terms of the sheet conductivity,

\begin{equation}\label{eq}
    \begin{aligned}
        \sigma(B)=-\mathcal{A}\frac{2e^2}{h}
        \Bigg[
        &\frac{3}{2}
        \left(
        \frac{1}{l\sb{\phi}^2}
        +\frac{4}{3\,l\sb{SO}^2}
        +\frac{1}{D\,\tau_B}
        \right)^{-1/2}\\
        &-\frac{1}{2}
        \left(
        \frac{1}{l\sb{\phi}^2}
        +\frac{1}{D\,\tau_B}
        \right)^{-1/2}
        \Bigg]+\sigma\sb{0},
    \end{aligned}
\end{equation}

\noindent where $l\sb{\phi}$ is the phase-coherence length, $l\sb{SO}$ is the spin-orbit length, $D$ is the diffusion constant, and $\tau\sb{B}=3\,l\sb{m}^4/W^2D$, with magnetic length $l\sb{m}=\sqrt{\hbar/eB}$ and wire width $W$. 
The prefactor $\mathcal{A}$, with units of inverse length, accounts for the effective geometry and conversion from the conductance correction of an individual wire to the sheet conductivity of the network, while $\sigma\sb{0}$ represents a field-independent background. 
We assume an effective channel width of $W=120$~nm, slightly narrower than the frame-gate electrode to account for electrostatic fringing fields. 
Using $W=150$~nm gives consistent trends, although with slightly worse agreement between the fits and measured curves.
The fitting parameters are $\sigma\sb{0}$, $\mathcal{A}$, $l\sb{\phi}$, and $l\sb{SO}$.

The fits are performed on the sheet conductivity, with the resulting curves converted back to resistivity for comparison with the experimental data. 
Before fitting, we correct for a residual field offset of $B\sb{offset}=-1.67$~mT, apply three-point smoothing, and restrict the fitting range to $\pm90$~mT. 
Measurements at several fixed temperatures yield consistent gate-dependent trends, with more stable fits obtained at $T=150$~mK, which we present in the main text. 
Since the model describes individual quasi-one-dimensional wires rather than an interconnected network, we focus on the relative evolution of $l\sb{SO}/l\sb{\phi}$ rather than interpreting the absolute fitted length scales. 
For $V\sb{FG}<-0.4$~V, the WAL feature becomes weak or unresolved, limiting the reliability of the extracted spin-orbit length in this regime.

\bibliography{Literature.bib}